# Crystal chemistry search of multiferroics with the stereochemically active lone pair

L. M. Volkova, D. V. Marinin

Institute of Chemistry, Far Eastern Branch of Russian Academy of Science, 159, 100-Let Prosp., Vladivostok 690022, Russia

**Abstract**

On the basis of our previous studies of magnetoelectric ordering of $BiFeO_3$, $TbMnO_3$, $TbMn_2O_5$ and $BiMn_2O_5$ we formulate the crystal chemistry criteria for the search of multiferroics and reveal potential multiferroics $Pb_2Cu(OH)_4Cl_2$, $Pb_5Cr_3F_{19}$, $Mn(SeO_3) \cdot H_2O$ and $BiPbSr_2MnO_6$ each containing the ion with a lone pair.

**Keywords** Multiferroic, Crystal chemistry search.

## 1 Introduction

It is generally known [1] that multiferroics [2] comprise a rather sparse class of materials. This is related to the fact that such materials must have both magnetic and electric ordering simultaneously in a single phase, while the interaction of the above ordered subsystems must have a linear character [3]. However, there exists a limitation on the coexistence of electric and magnetic orderings explained mainly by the conclusion that $d$ electrons of transition metals, which serve as a source of magnetism, reduce the possibility of the existence of ferroelectric distortions resulting into the loss of the symmetry center. The following phenomena are considered as an additional electronic or structural support of a simultaneous realization of ferromagnetic and ferroelectric transitions in a single phase: charge ordering in magnetic oxides [4], stereochemical activity of electron lone pairs [5, 6], and the Jahn-Teller distortion [7].

In search of potential novel multiferroics it is important to know how and why both the magnetic and the electric ordering emerge in a single phase. A correct solution of the problem of emerging of magnetoelectric ordering can be attained with using quantum chemistry methods based on the "first-principles". The accuracy and correctness of quantum chemistry calculations of the crystal structure, its stability and properties increase continuously. One of the examples here can be the Heyd-Scuseria-Ernzerhof hybrid functional in studying multiferroic systems such as $BiFeO_3$ and orthorhombic $HoMnO_3$ [8]. However, there are still difficulties in prediction of the compounds structure and properties from those "first-principles".

The crystal chemistry does not have answers to the above questions, however, it is capable to determine under which crystal chemical conditions this might happen and find the compounds with these conditions fulfilled.

In the present work we have performed the following studies:
- formulation the criteria and approaches to the search of novel multiferroics;
- search of novel potential multiferroics in the Inorganic Crystal Structure Database (ICSD) (version 1.4.6, FIZ Karlsruhe, Germany, 2010-2) among the stoichiometric compounds containing a magnetic ion and an ion with the stereochemically active lone pair of electrons using the calculation of magnetic interaction parameters by the crystal chemistry method [9, 10].

## 2 Crystal chemistry criteria for the search of multiferroic

From the crystal chemistry point of view, the common feature of the emerging of both the electrical polarization and the magnetic ordering is the ions displacement. By establishing the relation between the crystal chemistry parameters and the emerging of the magnetically induced ferroelectricity in $TbMnO_3$ [11], $TbMn_2O_5$ and $BiMn_2O_5$ [10], we determined three necessary conditions whose structural availability can produce magnetoelectric ordering under the effect of an external magnetic field. First, the existence of a competition of magnetic interactions on the specific geometric fragments in the paramagnetic phase. Second, the presence of intermediate ions in critical positions displacements from which might cause reorientation of the magnetic interactions between spins resulting in magnetic ordering. Third, the displacements accompanying the magnetic ordering should be polar (like in $TbMnO_3$) or create necessary conditions for emerging of polar displacements (for example, the charge disordering, like in $TbMn_2O_5$ and $BiMn_2O_5$) with preservation of the magnetic ordering. The fulfilment of three above conditions substantially reduces the probability of the multiferroics existence and explains, from the crystal chemistry point, "why are there so few magnetic ferroelectrics" [1].

The crystal chemistry studies of the room-temperature multiferroic $BiFeO_3$ [12] enabled us to correct the criteria of search of novel multiferroics having high temperatures of electrical and magnetic ordering. The main search criteria are the presence of the electrical polarization and the structurally conditioned magnetic ordering in the compounds of interest at room temperature. The existence of the competition of magnetic interactions is acceptable, if it can be eliminated by reorientation of the spins of the competing interactions (AFM – FM transition) through displacements of intermediate ions in critical positions. In the latter case, under the magnetic field effect, there is the possibility of the phase transition from the state with competing magnetic interactions to a uniform magnetic structure with preservation of the electrical ordering and increase of the electrical polarization due to emerging of the linear magnetoelectric effect. The search should be performed among the compounds with a stereochemically active lone pair or a magnetic ion with a directed distortion of the coordination polyhedron which can be induced by the Jahn-Teller effect or other reasons. Besides, it is possible to use the compounds containing the magnetic ion with an unstable (non-rigid) coordination surrounding.

## 3 Potential multiferroics

The search of potential multiferroics was performed in the Database ICSD among the stoichiometric compounds containing a magnetic ion and an ion with the stereochemically active lone pair of electrons $Tl^{1+}$, $Sn^{2+}$, $Pb^{2+}$, $Sb^{3+}$, $Bi^{3+}$, $Se^{4+}$ or $Te^{4+}$. The sign and strength of magnetic interactions in compounds were calculated by crystal chemical method [9, 10] on the basis of structural data with using the program "MagInter". As a result, we found 4 previously unstudied (according to our data) multiferroics. Regretfully, the composition of these compounds is not as simple as in the case of $BiFeO_3$, all the simple multiferroics with a lone pair included into the Database ICSD must have been already studied. Let us consider the tentative multiferroics.

3.1 $Pb_2Cu(OH)_4Cl_2$

In the crystal structure of diaboleite, $Pb_2Cu(OH)_4Cl_2$, (tetragonal, $P4mm$, $a$ = 5.880, $c$ = 5.500 Å, z = 1) [13] (Fig. 1a) the $Cu^{2+}(OH)_4Cl_2$ octahedra have the typical [4 + 2] Jahn-Teller distortion. Four $(OH)^-$ anions (d(Cu-O) = 1.972 Å) occupy the equatorial positions and two $Cl^-$ anions (d(Cu-Cl1) = 2.551 and 2.949 Å) occupy the apical positions of the octahedron at short (d(Cu-Cl1) = 2.55 Å) and long (d(Cu-Cl1) = 2.95 Å) distances. The $Cu^{2+}$ cations are displaced by 0.34 Å from the equatorial anions plane in the direction of the nearer apical Cl atom. The coordination surrounding of the $Pb^{2+}$ ion containing the stereochemically active lone pair of electrons includes, from one side, four OH groups at short distances (d(Pb-O) = 2.46 Å) and, from another side, four Cl ions at long distances Cl (d(Pb-Cl1) = 3.22 Å X2 and d(Pb-Cl2) = 3.40 Å X2). Orientation of the long axial bonds Cu-Cl1 of the $Cu^{2+}$ octahedra and the electronic density of the $Pb^{2+}$ ion lone pairs along the $c$ axis induces electrical polarization in this very direction. The $Cu(OH)_4Cl_2$ octahedra form corner-sharing chains along the $c$ axis, and these chains are cross-linked by $Pb^{2+}$ cations and hydrogen bonding.

In the linear chains along the $c$-axis, there exist the strong AFM nearest-neighbor $J1_1$ ($J1_1$ = -0.1162 Å$^{-1}$, d(Cu-Cu) = 5.500 Å = c) couplings (Fig. 1b). The contribution to the AFM component of this couplings results from the apical oxygen ion Cl, which is located on a direct line connecting Cu ions. The dominant intrachain AFM $J1_1$ couplings determine the structure of the magnetic system of this compound as a strong one-dimensional antiferromagnet, since there is no competition between the $J1_1$ couplings and the FM next-nearest-neighbor $J1_2$ ($J1_2/J1_1$ = -0.13, (d(Cu-Cu) = 11.00) couplings in a chain and the interchain couplings are weak. The strongest of the interchain couplings FM $J2_1$ ($J2_1/J1_1$= -0.23, d(Cu-Cu) = 5.880 Å) emerging along the $a$ and $b$ axes are 4.4-fold weaker than the intrachain nearest-neighbor couplings. The FM $J2_1$, AFM $J3$ ($J3/J1_1$ = 0.07, d(Cu-Cu) = 8.051 Å), FM $J4$ ($J4/J1_1$ = -0.005, d(Cu-Cu) = 8.316 Å) and AFM $J5$ ($J5/J1_1$= 0.22, d(Cu-Cu) = 9.970 Å) interchain couplings do not compete with the AFM $J1_1$ couplings in the triangles $J1_1J2_1J3$ and $J1_1J4J5$. Nevertheless, there is still a competition in the planes perpendicular to the $c$ axis. The next-nearest-neighbor AFM $J2_2$ ($J2_2/J1_1$= 0.06, d(Cu-Cu) = 11.760 Å) couplings compete with the nearest-neighbor FM $J2$ ($J2_2/J2_1 = -0.27$) couplings along the $a$ and $b$ axes and with very weak FM $J4$ couplings in the triangles $J2_2J4J4$. However, slight displacements of the O1 ions could induce the transition of the $J2_2$ couplings from the AFM to the FM state and eliminate the above competition at preserving the electrical polarization.

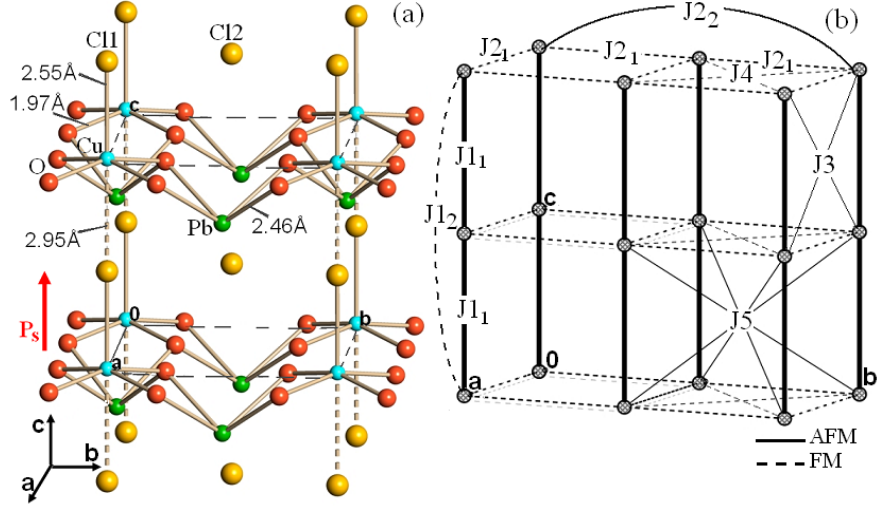

**Fig. 1** The crystal structure of $Pb_2Cu(OH)_4Cl_2$ (a) and the sublattice of $Cu^{2+}$ and coupling the $J_n$ (b). In this and other figures the thickness of lines shows the strength of $J_n$ coupling. AFM and FM couplings are indicated by solid and dashed lines, respectively.

One can suggest two models of the centrosymmetrical paraelectric phase in the tetragonal P4/mmm (N123) space group. In the model I ($a \sim 5.885$, $c \sim 5.505$ Å, z = 1) two Pb ions are disordered in the position 4i. The ions Cu, O, Cl1 and Cl2 occupy the positions 1a, 4j (x = 0.238, y = 0.238), 1b and 1d, respectively. In the model II ($a \sim 5.885$, $c \sim 11.010$ Å, z = 2) the parameter $c$ is doubled and one can observe alternating layers with opposite orientation of the lone pairs of $PbO_4E$-groups and shortened Cu-Cl bonds. The ions occupy the following positions: Pb1 – 2g, Pb2 – 2e, Cu – 2h (x = 0.5, y = 0.5, z = 0.275), O – 8r (x = 0.264, y = 0.264, z = 0.271), Cl1 - 1c, Cl2 - 1d and Cl3 - 2f.

3.2 $Pb_5Cr_3F_{19}$

The compound $Pb_5Cr_3F_{19}$ (tetragonal I4cm(N108): $a$ = 14.384 Å, $c$ = 7.408 Å at 22º C) [14] (Fig. 2a) is a known ferroelectric with the ferroelectric-paraelectric transition at $T_c$ = 282º C. Its crystalline structure consists of individual $Cr1F_6^{3-}$ and corner-sharing chain-forming $Cr2F_6^{3-}$ octahedra and two types of distorted Pb polyhedra with the stereochemically active lone pair of electrons. It was shown in [14] that the Cr ions were significantly displaced along the polar direction of the $c$ axis from the centers of the $Cr1F_6$ and $Cr2F_6$ octahedra. The Cr1 ion is linked through short bridge bonds Cr-F-Pb with four Pb1 ions (through two F1 ions and one F4 ion) and one Pb2 ion (through the F6 ion), while the Cr2 ion is linked only with four Pb1 ions through 4 F3 ions located in the octahedron equatorial plane.

According to our calculations, the nearest AFM $J1_1$ ($J1_1 = -0.192$ Å$^{-1}$, d(Cr2-Cr2) = 3.704 Å) couplings in the linear chains of the Cr2 ions are dominating (Fig. 2b). The next-nearest neighbors couplings $J1_2$ ($J1_2/J1_1 = -0.20$, d(Cr2-Cr2)=7.408 Å) in the chains are of the FM type and do not compete with the $J1_1$ couplings. A weak competition is anyway present in the chain, since the couplings with the third neighbors $J1_3$ ($J1_3/J1_1 = -0.10$, d(Cr2-Cr2)=11.112 Å) are also of the FM type; however, they can easily transform into the AFM type at slight displacement of the F3 ions to the Cr2-Cr2 bond line and stop the competition. Such a displacement of the F3 ions would not change the sign of the $J1_1$ and $J1_2$ couplings. The strongest $J9$ ($J9/J1_1 = 0.14$, d(Cr1-Cr1)=7.613 Å) coupling between the Cr1 ions is of the AFM type and seven-fold weaker than the $J1_1$ coupling in the Cr2 ion chain. The remaining AFM couplings (J5: $J5/J1_1 = 0.04$, d(Cr1-Cr1)=5.980 Å; J6: $J6/J1_1 = 0.07$, d(Cr1-Cr1)=6.640 Å; J10: $J10/J1_1 = 0.08$, d(Cr1-Cr1)=10.767 Å) and FM (J2: $J2/J1_1 = -0.006$, d(Cr1-Cr1)=5.118 Å; J7: $J7/J1_1 = -0.03$, d(Cr1-Cr1)=7.408 Å) between the Cr1 ions are even weaker. The AFM J4 ($J4/J1_1 = 0.08$, d(Cr2-Cr1)=5.851 Å) and FM J3 ($J3/J1_1 = -0.03$, d(Cr2-Cr1)=5.565 Å) and J8 ($J8/J1_1$ = -0.05, d(Cr2-Cr1)=7.427 Å) couplings between the Cr2 ion chain and the Cr1 ions are also very weak.

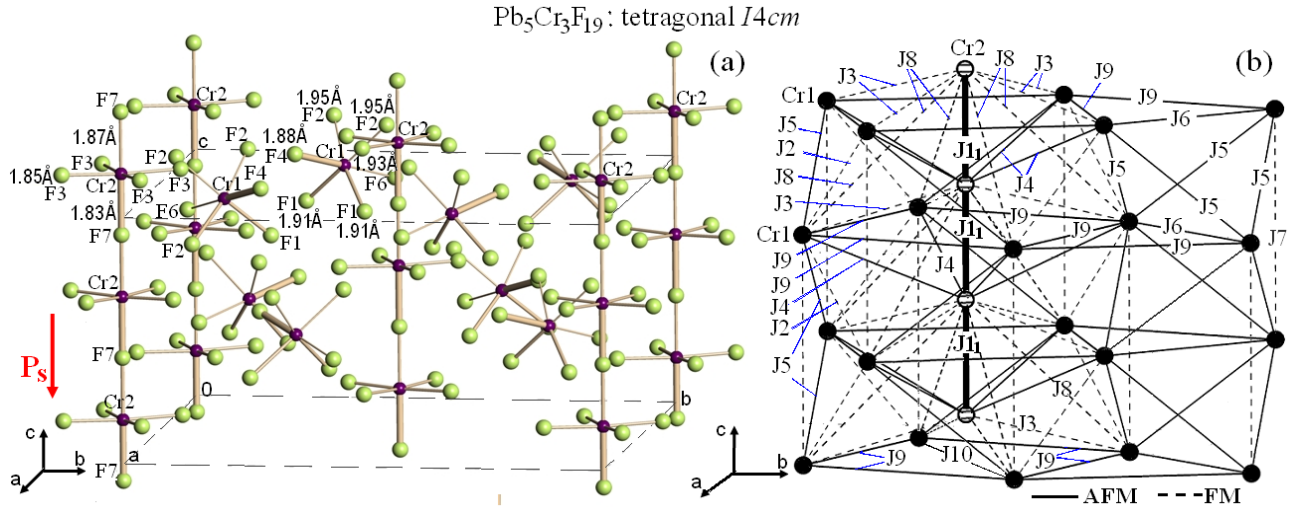

**Fig. 2** The relation of Cr–F bond lengths in Cr1 and Cr2 octahedra in the crystal structure of $Pb_5Cr_3F_{19}$ (a) and the coupling $J_n$ (b). In this and other figures the thick and thin lines refer to short and long M-X bonds, respectively.

The magnetic structure of $Pb_5Cr_3F_{19}$ at a temperature 22º C must be non-collinear, since there exists, aside from a small competition inside the chains, the competition between other magnetic compounds (Fig. 2b). The AFM $J9$ and FM $J3$ couplings compete with other couplings in the following triangles formed by one Cr2 ion and two Cr1 ions: AFM $J9$ - AFM $J9$- AFM $J6$, AFM $J9$ - AFM $J9$- AFM $J10$, AFM $J9$ - AFM $J4$ - AFM $J4$, AFM $J9$ - FM $J3$- FM $J3$, AFM $J9$ - FM $J8$ - FM $J8$, FM $J3$ - FM $J2$ - FM $J4$ and FM $J3$ - FM $J8$ - AFM $J5$. Besides, there exists a competition in the triangle FM J3 - FM J8 - AFM $J1_1$ formed by two Cr2 ions and one Cr1 ion and in the triangle AFM J6 - AFM J5 - AFM J5 formed by Cr1 ions. Elimination of competition in the above triangles might take place, for example, at transition of the AFM $J9$, $J6$ and $J10$ couplings into the FM state and FM $J3$ coupling – into the AFM state during displacement of the Pb1 and F2 ions for the $J9$ coupling, the F2 and F6 ions for the $J6$ coupling and the F3 ions for the $J10$ and $J3$ couplings. One can assume that the change of the stereochemical activity of the $Pb^{2+}$ lone pair under the temperature effect or external magnetic field imposing could induce the above displacements. For example, in the ferroelectric $Pb_3Al_3F_{19}$ [15], whose crystalline structure at -113º C corresponds to that of the ferroelectric $Pb_5Cr_3F_{19}$ at 22º, the temperature impact changes the orientation and tilts of the individual and the corner-sharing chain-forming $AlF_6$ octahedra as well as the Cr-F bond lengths in these octahedra.

It is complicated to conclude on the temperature of phase transition in the ferroelectric $Pb_5Cr_3F_{19}$ from the space-modulated to the homogeneous antiferromagnetic structure on the basis of structural data at a single temperature. One should perform studies of the magnetic properties changes under the magnetic field effect in the course of temperature reduction, starting from 282º C.

We assume that the possible structure of the paraelectric phase for $Pb_5Cr_3F_{19}$ can belong to one of two space groups: $I4/m$ (N87) and $I4/mcm$ (N140). The crystal structures of $Pb_5Cr_3F_{19}$ are not experimentally determined in these space groups. For the calculations one can use the atom coordinates of the isostructural compound $Pb_5Al_3F_{19}$ in the space group I4/m determined in [16] using the X-ray diffraction from single crystal and the coordinates of the theoretical model of $Pb_5Cr_3F_{19}$ in the space group I4/mcm suggested in [15].

3.3 $Mn(SeO_3) \cdot H_2O$

In the crystal structure of the non-centrosymmetrical orthorhombic polymorph $Mn(SeO_3) \cdot H_2O$ ($Ama2$: $a$ = 5.817 Å, $b$ = 13.449 Å, $c$ = 4.8765 Å) [17] (Fig. 3a and b), the octahedra $MnO_6$ are coupled by common vertices O3 into corrugated perovskite-like layers in the (010) plane. The $MnO_6$ octahedra layers are formed from $SeO_3$ trigonal pyramids whose stereochemically active lone pairs are directed between layers. The manganese ions are displaced (by 0.26 Å) along the $c$ axis from the octahedral centers to their facets. The difference in Mn-O distances to two opposite facets is in the range from 0.08 up to 0.15 Å. The latter indicates to the electric polarization $P_s$ along the direction [00$\bar{1}$].

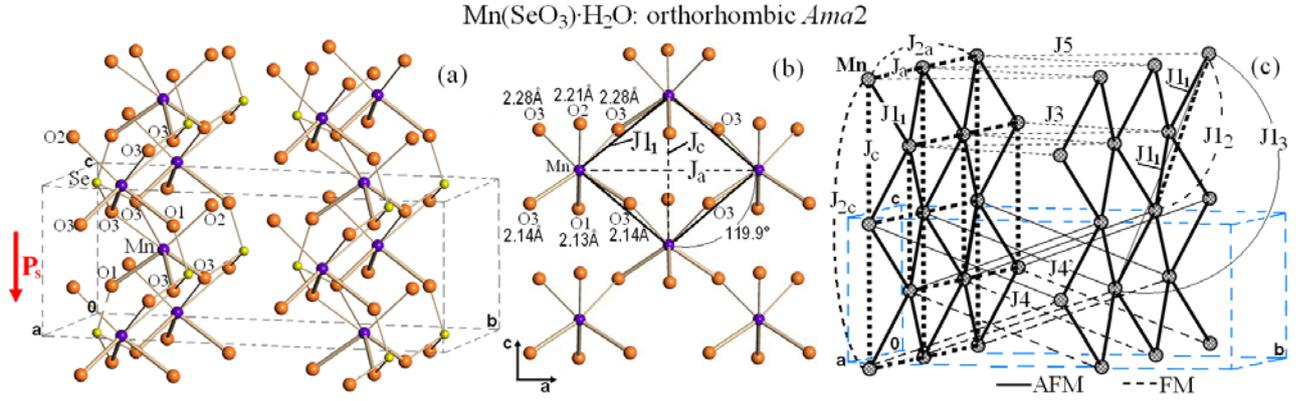

**Fig. 3** The crystal structure of the orthorhombic polymorph Mn(SeO$_3$)·H$_2$O (a) and corrugated perovskite-like layers in the (010) plane (b) and sublattice of Mn$^{2+}$ and the coupling $J_n$ (c).

The magnetic ions Mn$^{2+}$ form a corrugated "square" lattice in the (010) plane (Fig 3c). The distorted "squares" sides are identical, while the angles differ (79.22° and 99.02°). According to our calculations, there is no competition of magnetic couplings in the above lattice, since the nearest-neighbor $J1_1$ ($J1_1$ = -0.0112 Å$^{-1}$, d(Mn-Mn) = 3.824 Å) couplings along the "square" side are of the AFM type, while the second $J_c$ ($J_c/J1_1$ = -3.018, d(Mn-Mn) = 4.877 Å) and $J_a$ ($J_a/J1_1$ = -2.518, d(Mn-Mn) = 5.817 Å) couplings along diagonals of the "square" are of the FM type (Fig. 3c). Besides, there is no competition between the nearest-neighbor AFM $J1_1$ couplings and the next-nearest neighbor FM $J1_2$ ($J1_2/J1_1$ = -1.51, d(Mn-Mn) = 7.591 Å) and AFM $J1_3$ ($J1_3/J1_1$ = 0.14, d(Mn-Mn) = 11.396 Å) couplings along the square lattice sides. However, there is a competition between the FM $J3$ ($J3/J1_1$ = -0.42, d(Mn-Mn) = 6.897 Å), FM $J4$ ($J4/J1_1$ = -0.68, d(Mn-Mn) = 7.153 Å) and FM $J5$ ($J5/J1_1$ = -0.09, d(Mn-Mn) = 7.761 Å) interlayer couplings and the AFM $J1_1$ intralayer couplings in the triangles $J1_1J4J3$ and $J1_1J4J5$. One should emphasize that the peculiarities of intermediate ions locations in local coupling spaces in the structure Mn(SeO$_3$)·H$_2$O create the situation when slight displacements of SeO$_3$ groups and H$_2$O might produce substantial strengthening of the AFM $J1_1$ couplings up to -0.0288 Å$^{-1}$ and weakening and/or AFM-FM transitions of interlayer couplings and elimination of the above competition.

We have calculated the magnetic couplings parameters in another polymorph Mn(SeO$_3$)·H$_2$O, which is centrosymmetrical (monoclinic $P2_1/n$ space group) [17], does not have the electrical polarization and differs from the former one by the stacking of the layers along the $b$-axis. The structural changes result in non-equivalency of the nearest neighbor couplings in the distorted square lattice. One of them $J1'_1$ ($J1'_1$ = -0.0305 Å$^{-1}$, d(Mn-Mn) = 3.841 Å) is of the AFM type and 4.2-fold stronger than another FM $J1_1$ coupling. The parameters of the couplings along the square diagonal do not virtually change. As a result, the square lattice in the monoclinic polymorph is frustrated because of the competition in the triangles $J1_1$(FM)-$J1'_1$(AFM)-$J_a$(FM) and $J1_1$(FM)-$J1'_1$(AFM)-$J_a$(FM) and $J1_1$(FM)-$J1'_1$(AFM)-$J_c$(FM), where $J1'_1$ = -4,24$J1$ = -0.86$J_a$ = -0.98$J_c$.

To sum up, the orthorhombic polymorph Mn(SeO$_3$)·H$_2$O, unlike the monoclinic polymorph, contains AFM corrugated "square" lattices without competition of magnetic couplings and can be described as ferroelectric (electrical polarization along the $c$-axis).

3.4 BiPbSr$_2$MnO$_6$

The emerging of the magnetically induced ferroelectricity is possible in the layered oxide BiPbSr$_2$MnO$_6$. There are two known variants of the crystal structure of the mentioned compound. According to the single crystal X-ray diffraction studies, the crystal structure of BiPbSr$_2$MnO$_6$ is noncentrosymmetrical (orthorhombic space group $A2aa$: $a$ = 5.331Å, $b$ = 5.399 Å, $c$ = 23.751 Å) [18] (Fig. 4a), while the powder neutron diffraction studies demonstrate that the above structure is centrosymmetrical (orthorhombic space group $Amaa$: $a$ = 5.320Å, $b$ = 5.380 Å, $c$ =23.714) [19]. The atomic parameters $x, y, z$ are virtually identical in both structural variants. Their main difference consists in the fact that for the centrosymmetrical structure the oxygen atoms in the (Bi,Pb)O layer are disordered over two positions along the $a$-axis. The octahedra MnO$_6$ are coupled by common vertices O1 into perovskite-type layers in the $ab$ plane.

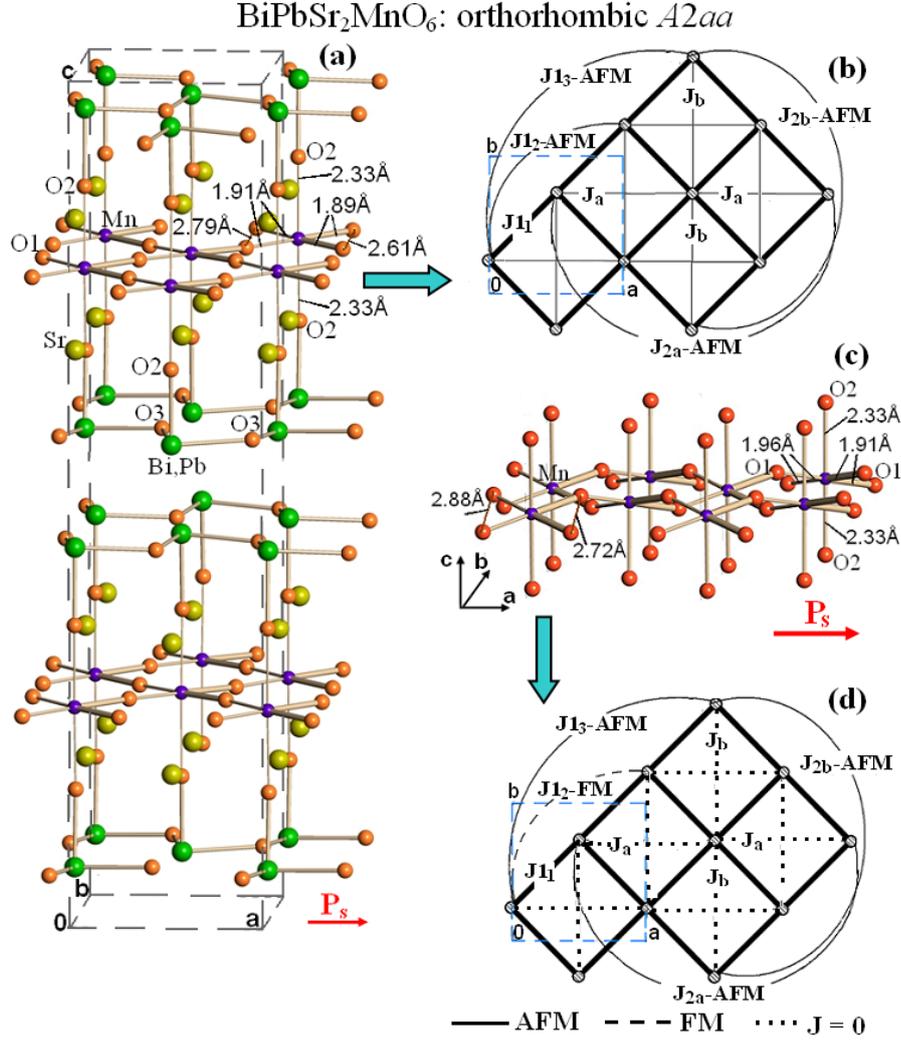

**Fig. 4** Crystal structure of BiPbSr$_2$MnO$_6$ in space group *A2aa* [18] (a) and the sublattice of Mn$^{3+}$ and competing couplings $J_n$ in MnO$_2$ plane (b). Ferroelectric model perovskite-type layers at displacements of the ion O1 by 0.33 Å along the direction $00\bar{1}$ and the ion Mn by 0.02 Å along the direction 100 (c) and the magnetically ordered $J_n$ coupling in the model (d).

In the noncentrosymmetrical structure the manganese ion is displaced (by 0.06 Å) from the octahedron center along the *a* axis in the direction $\bar{1}00$ to the longest edge (d(O1-O1)=2.79 Å) in the equatorial plane. However, in spite of this displacement, the distances (d(Mn-O1) = 1.91 Å) to the oxygen ions forming this edge remain longer (by 0.02 Å) than those (d(Mn-O1) = 1.89 Å) to the same ions forming the opposite octahedron edge, since the latter is significantly shorter (d(O1-O1)=2.61 Å).

In both of the above structures the ions Bi$^{3+}$ and Pb$^{2+}$ equally share the same position. With two nearest oxygen ions, one of which serves as the axial vertex of the MnO$_6$ octahedron, they form tetrahedra (Bi,Pb)O$_3$E, thus attaching to the manganese octahedral layers from both sides. The stereochemically active lone electron pairs of Bi$^{3+}$ and Pb$^2$ do not polarize the structure, since their directions are different: they are directed to the space between formed sandwich-like layers, thus separating them from each other.

The respective magnetic couplings in noncentrosymmetrical (I) and centrosymmetrical (II) structures are virtually identical. The crystal sublattice of the magnetic ions Mn$^{3+}$ comprises a slightly distorted square lattice in the plane *ab* with the square side d(Mn-Mn)=3.794(3.783) Å and the angles 90.73º(90.63º) and 89.27º(89.37º) in the structures I(II) (Fig. 4b). The magnetic couplings in perovskite layers emerge only under effect of intermediate ions O1. The ions O2 are not included into the bounded space region between magnetic ions Mn due to the Jahn-Teller elongation of the bonds Mn-O2 (d(Mn-O2) = 2.327x2 Å) along the axis *c*.

The dominating AFM nearest neighbor $J1$ ($J1$ = -0.184(185) Å$^{-1}$ in I(II)) couplings along the square sides are initiated by the bridge oxygen ions O1 located virtually in the middle of the bond line Mn-Mn (angle Mn-O1-Mn = 175.97º(175.69°)). The AFM $J1$ couplings in the lattices are 18-49-fold stronger than the AFM $J_c$ ($J_c/J1$ = 0.03(0.05), d(Mn-Mn) = 12.171(12.152) Å) and $J'_c$ ($J'_c/J1$ = 0.05(0.02), d(Mn-Mn) = 12.178(12.158) Å) couplings between the lattices.

In the square lattices, there exists a weak competition, which could be the reason of the weak ferromagnetism found in [20]. First, strong AFM $J1$ couplings compete with very weak AFM second $J_a$ ($J_a/J1$ = 0.013(0.013), d(Mn-Mn) = 5.331(5.320) Å) and $J_b$ ($J_b/J1$ = 0.017(0.017), d(Mn-Mn) = 5.399(5.380) Å) couplings along the diagonal of the square and $J1_2$ ($J1_2/J1$ = 0.017(0.019), d(Mn-Mn) = 7.587(7.566) Å) couplings with the second neighbors along the square lattice sides. Second, weak AFM $J_a$ couplings compete with stronger AFM $J_{2a}$ ($J_{2a}/J_a$ = 6.5(6.4), d(Mn-Mn) = 10.662(10.641) Å) and $J_{2b}$ ($J_{2b}/J_b$ = 5.3(5.4), d(Mn-Mn) = 10.798(10.759) Å) couplings with the second neighbors along the square lattice diagonals. However, the perovskite structures are characterized by the location of a part of intermediate ions in the critical positions 'a' and 'b'[10] in local spaces of $J1_2$ and $J_a(J_b)$ couplings, respectively. During the displacement of intermediate ions from the above positions, it is possible that the sign and value of their contribution to the emerging of magnetic interactions will change, thus inducing the AFM–FM transition and/or dramatic change of the magnetic interaction strength. The performed calculations demonstrate that the ion O1 displacement by 0.33 Å along the direction $00\bar{1}$ (the ion O1 coordinate change from z = 0.2483 to z = 0.2343) within the space group $A2aa$ would result in the emerging of magnetic ordering due to the processes described below (Fig. 4c and d),. There will take place the transition of the AFM $J1_2$ coupling into the FM state and, therefore, it will stop competing with the AFM $J1$ coupling. The AFM $J_a$ and $J_b$ couplings ($J_a$ = 0, $J_b$ = 0) and their competition with the AFM $J_{2a}$ and $J_{2b}$ couplings will disappear. Then one can make a reverse conclusion that the emerging of the magnetic ordering under magnetic field effect could induce the displacement of the ion O1 by 0.33 Å along the direction $00\bar{1}$. The distortion of the perovskite layer induced by this displacement is possible due to the absence of rigid bonds along the axes $b$ and $c$ between Bi(Pb)O$_3$ groups because of stereoactivity of the lone pairs. The increase of electrical polarization along the direction 100 at preservation of magnetic ordering can occur as a result of the displacement of Mn ions along the same direction.

To sum up, for the compound BiPbSr$_2$MnO$_6$ under the magnetic field effect, we demonstrated the possibility of the phase transition from the state with competing magnetic interactions into a homogeneous magnetic structure to be accompanied with the emerging of electrical polarization.

## 4. Conclusions

On the basis of our previous the results of studies of the role of crystal chemistry factors in emerging of magnetoelectric properties in BiFeO$_3$, TbMnO$_3$, TbMn$_2$O$_5$ and BiMn$_2$O$_5$, we determined main criteria for the search of novel multiferroics and found 4 potential compounds of this type. These compounds have the electrical ordering and, according to our calculations, the magnetic ordering at room temperature or can transform into the magnetically ordered state under effect of the applied magnetic field.

## Acknowledgments

This work is supported by grant 09-I-P18-03 of the Far Eastern Branch of the Russian Academy of Sciences.